\documentclass{svmult}

\usepackage{makeidx}         
\usepackage{graphicx}        
\usepackage{multicol}        
\usepackage[bottom]{footmisc}

\makeindex             

\begin{document}

\title*{Blockbusters, Bombs and Sleepers:\\
{\Large The income distribution of movies}}
\titlerunning{The income distribution of movies} 
\author{Sitabhra Sinha and Raj Kumar Pan}
\institute{The Institute of Mathematical Sciences, C. I. T. Campus,
Taramani,\\ 
Chennai - 600 113, India.\\
\texttt{sitabhra@imsc.res.in}}

\maketitle
\noindent
The distribution of gross earnings of movies released each year show a 
distribution having a power-law 
tail with Pareto exponent $\alpha \simeq 2$. 
While this offers interesting parallels with 
income 
distributions of individuals, it is also clear that it cannot be 
explained by simple asset exchange models, as movies do not interact 
with each other directly. In fact, movies (because of the large quantity 
of data available on their earnings) provide the best entry-point for 
studying the dynamics of how ``a hit is born'' and the resulting 
distribution of popularity (of products or ideas). In this paper, we
show evidence of Pareto law for movie income, as well as, an analysis
of the time-evolution of income.

\vspace{-0.5cm}
\section{Introduction}
\label{sec:1}
\vspace{-0.25cm}
While the personal income distribution has been a subject
of study for a long time \cite{Par97}, it is only recently that other kinds
of income distribution, e.g., the income of companies \cite{Oku99}, 
have come under close scrutiny. More than a century ago, Vilfredo Pareto
had reported that the income distribution of individuals or households
follows a power law with an universal exponent of $\alpha = 1.5$. 
While recent studies
have shown this claim about universality to be untenable, it has indeed
been extensively verified that the higher-end (i.e., the tail) of the income, 
as well as wealth, distribution follows a power law. Whether similar power
laws occur for other types of income distribution is therefore of
high topical interest. 

\noindent
The income (or gross) of movies released commercially in theaters every year  
provides an opportunity to
study a very different kind of income distribution from those usually studied.
Not only is movie income a very well-defined quantity, but high-quality 
data is publicly available
from web-sites such as {\em The Numbers} \cite{numbers} and {\em Movie Times}
\cite{movietimes}. The income distribution, as well, as the time evolution
of the income, can be empirically determined with high accuracy. 
Movie income distribution is also of theoretical interest because such
a distribution clearly cannot be explained in terms of asset exchange models,
one of the more popular class of models used for explaining the nature
of personal income distribution. As movies don't exchange anything between
themselves, one needs a different theoretical framework to explain the 
observed distribution for movie income \cite{Sin04b}.

\noindent
Even more significantly, movie income can be considered to be a measure of
popularity \cite{Sin04}. 
Seen in this light, this distribution is a prominent member of 
the class of popularity distributions, that looks at how the success of
various products (or ideas) in appealing to public taste is distributed. 
Examples of such distributions include the popularity of scientific
papers as measured by the number of citations \cite{Red98}, books as measured
by the sales figures from an online bookstore \cite{Sor04}, 
etc. Of course, income is not the only measure
of a movies' popularity; e.g., one possibility is to use the number of votes
per film from registered users of IMDB \cite{imdb}. However, such voting
may not reflect the true popularity of movies as it costs nothing to 
give a vote. On the other hand, when one is voting with one's wallet, by
going to see a movie in a theater, it is a far more reliable indicator
of the film's popularity.

\vspace{-0.5cm}
\section{A Pareto Law for Movies}
\label{sec:2}
\vspace{-0.25cm}
Previous studies of movie income distribution \cite{Sor99,Dev99,Dev03}
had looked at limited data sets and found some evidence for a power-law
fit. A more rigorous demonstration has been given in Ref. \cite{Sin04},
where data for all movies released in theaters across USA during 1997-2003
were analysed. It was shown that the rank distribution of the opening gross 
as well as the total gross of the highest earning movies for all these years 
follow a power-law with an
exponent close to $-1/2$. As the rank distribution exponent is simply
the inverse of the cumulative gross distribution exponent
\cite{Red98}, this 
gives a power-law tail for the income distribution with a Pareto exponent
$\alpha \simeq 2$. It is very interesting that this value is identical
to that of corresponding 
exponents for citations of scientific papers \cite{Red98} and book sales
\cite{Sor04}, and is suggestive of an universal exponent for many different
popularity distributions.

\begin{figure}[tbp]
\centering
\includegraphics[width=0.48\linewidth,clip]{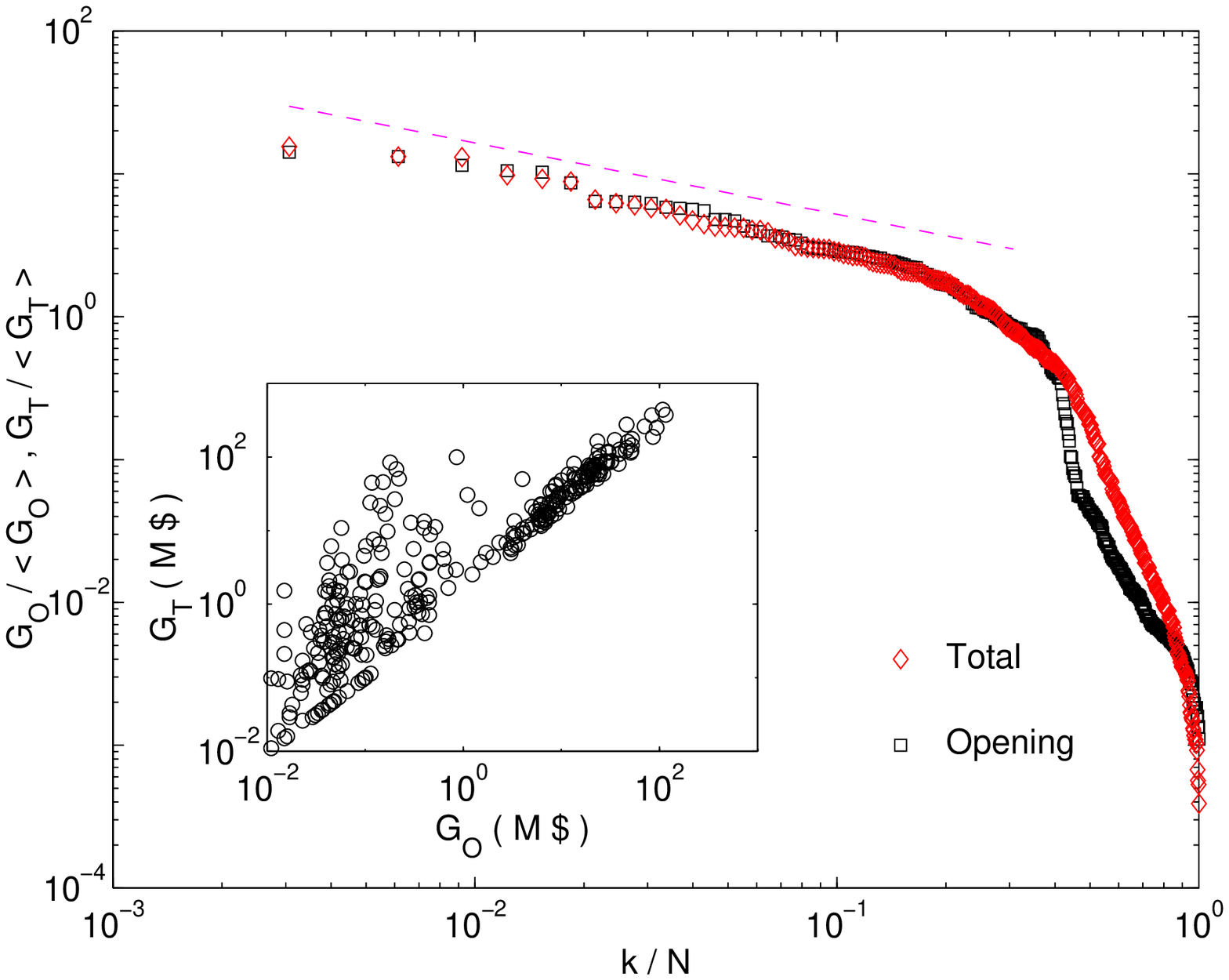}
\includegraphics[width=0.51\linewidth,clip]{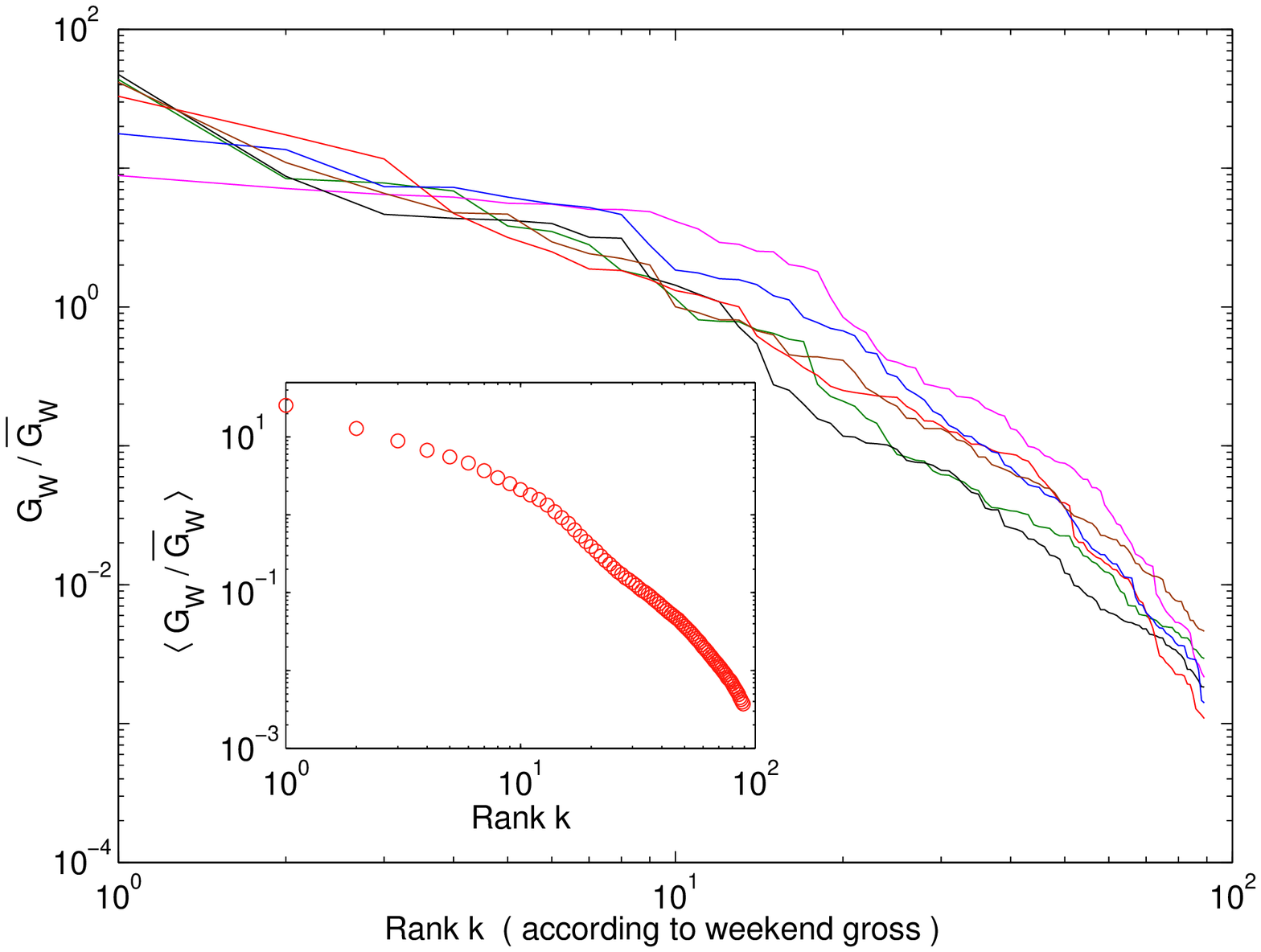}
\caption{Income distribution of movies released in theaters across USA
for the year 2004: (Left) Scaled rank-ordered plot of movies 
according to opening gross (squares) and total gross (diamonds). The
rank $k$ has been scaled by the total number of movies released that
year ($N = 326$) while the gross ($G_O, G_T$) has been scaled by
its average. The broken line of slope $-0.5$ has been shown for visual
reference. The inset shows the total gross earned by a movie, plotted
against its opening gross (in millions of \$). As indicated by the
data, there is a high degree of correlation between the two.
(Right) Scaled rank-ordered plot of movies according to weekend gross, $G_W$,
for six arbitrarily chosen weekends. The top 89 movies in a weekend are shown,
and the weekend gross of each movie has been scaled by the average weekend
gross of all movies playing that weekend. The inset shows the average of
the scaled rank-ordered plots for all the weekends in 2004.}
\label{fig:1}       
\vspace{-0.1cm}
\end{figure}
\noindent
Fig. \ref{fig:1} (left) demonstrates the Pareto law of movie income for the 
movies released across theaters in USA in 2004. 
Both the opening gross, $G_O$, as well as the total gross, $G_T$, 
(scaled by their respective
averages over all the movies released that year) show a power-law behavior
with the same exponent. The similarity of these two curves can be
partially explained from the inset figure, which shows that there is strong
degree of correlation between the income of a movie at its opening, 
and its total income. Movies which open poorly but perform well later 
({\em sleepers}) are relatively uncommon and are seen as the points deviating
from the linear trend in the inset figure.
Arguably, a better comparison with the Pareto distribution of personal income 
can be made by
looking at the income distribution of movies running on a particular
weekend [Fig. \ref{fig:1} (right)]. However, the smaller number of data
points available for such a plot means that the scatter is larger. As a result,
it is difficult to make a judgement on the nature of the weekend income
distribution. 

\vspace{-0.5cm}
\section{Time-evolution of movie income}
\label{sec:3}
\vspace{-0.25cm}
In this section, we focus on how the gross of a movie changes with time
after its theatrical release, until it is withdrawn from circulation.
Based on how they perform over this time, movies can be classified into
{\em blockbusters} having both high opening and high total gross, {\em bombs}
(or flops) having low opening as well as low total gross and {\em sleepers}
that have low opening but high total gross. Not surprisingly, the
corresponding theatrical lifespans also tend to be high to intermediate for
blockbusters, low for bombs and high to very high for sleepers.
  
\begin{figure}[tbp]
\centering
\includegraphics[width=0.48\linewidth,clip]{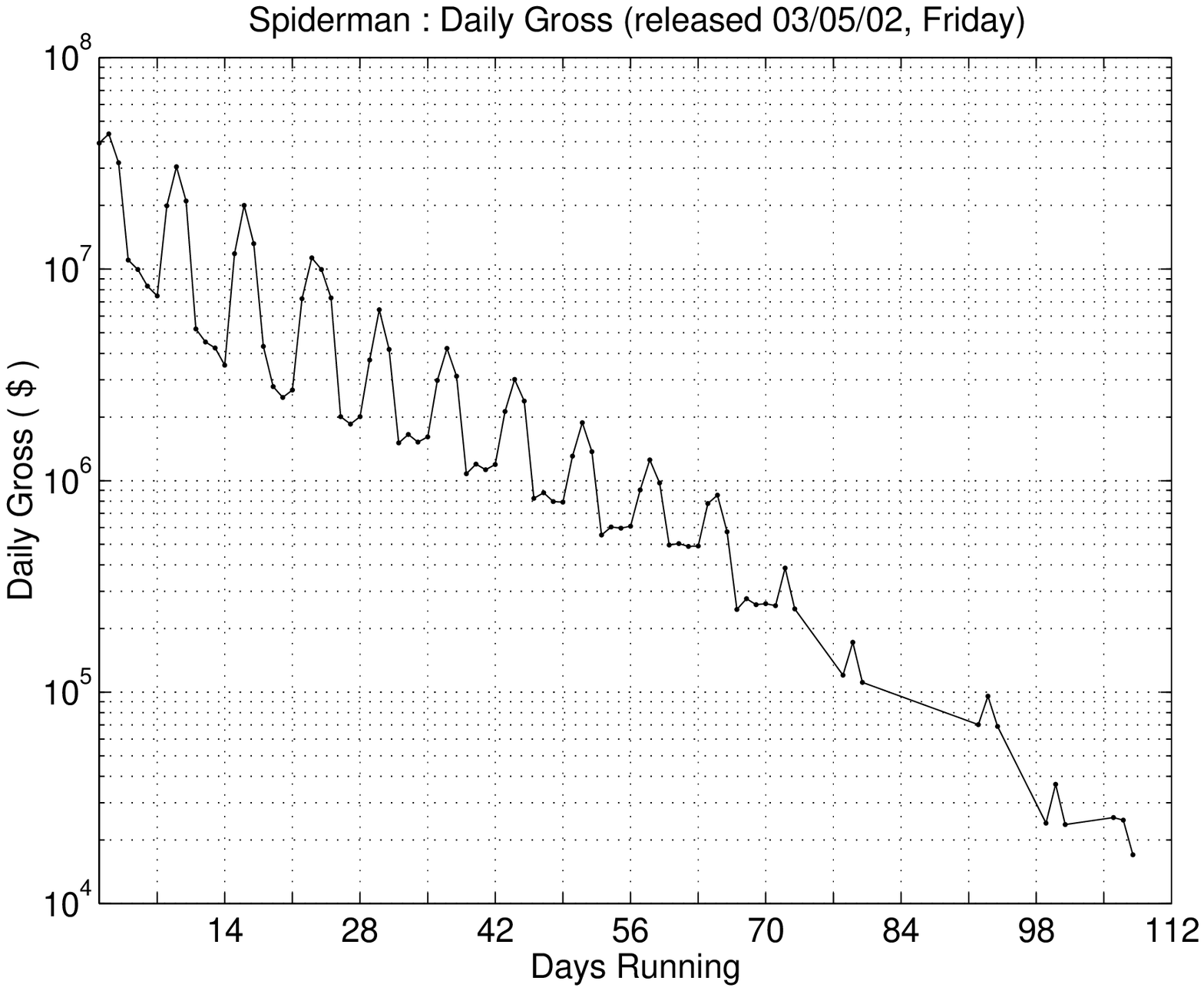}
\includegraphics[width=0.48\linewidth,clip]{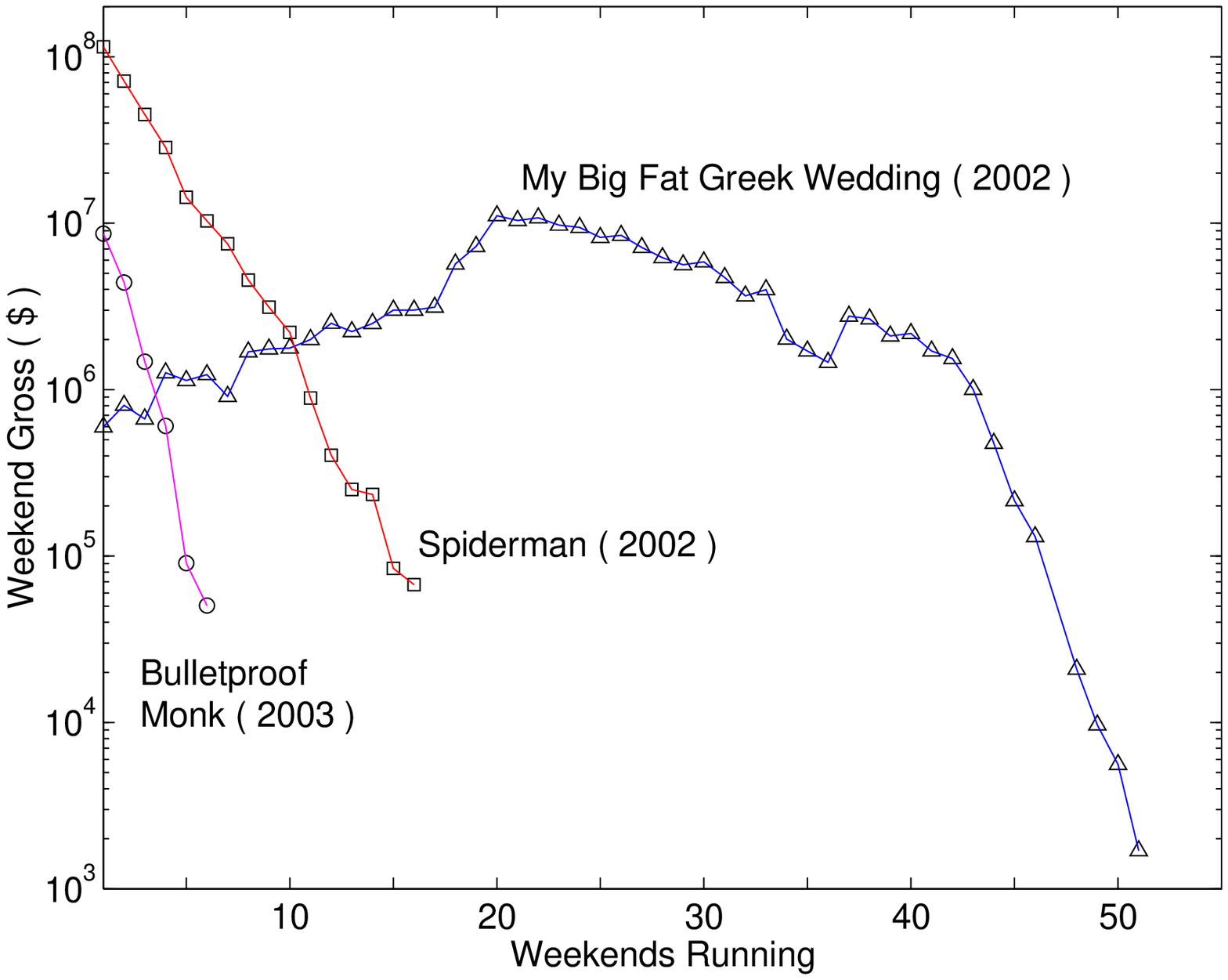}\\
\caption{ Classifying movies according to time-evolution of the gross (income):
(Left) Daily gross of a typical blockbuster movie (Spiderman) showing
weekly periodic fluctuations (with gross peaking on weekends), 
while the overall trend is
exponential decay. (Right) Comparing examples of blockbusters 
(Spiderman), bombs (Bulletproof Monk) and sleepers (My Big Fat Greek Wedding)
in terms of the time-evolution of weekend gross. 
Time is measured in weekends to remove intra-week fluctuations.
}
\label{fig:2}       
\vspace{-0.15cm}
\end{figure}

\noindent
Consider a classic blockbuster movie, {\em Spiderman} (released in 2002).
Fig. \ref{fig:2} (left) shows how the daily gross decays with time after
release, with regularly spaced peaks corresponding to large audiences
on weekends. To remove the intra-week fluctuations and observe the overall 
trend, we focus on the time series of weekend gross. This shows an exponential
decay, a feature seen
not only for almost all other blockbusters,
but for bombs as well [Fig. \ref{fig:2} (right)]. The only difference
between blockbusters and bombs is in their initial, or opening, 
gross. However, sleepers
behave very differently, showing an increase in their weekend gross and
reaching their peak performance (in terms of income) quite a few weeks after
release, before undergoing an exponential decay. 

\begin{figure}[tbp]
\centering
\includegraphics[width=0.48\linewidth,clip]{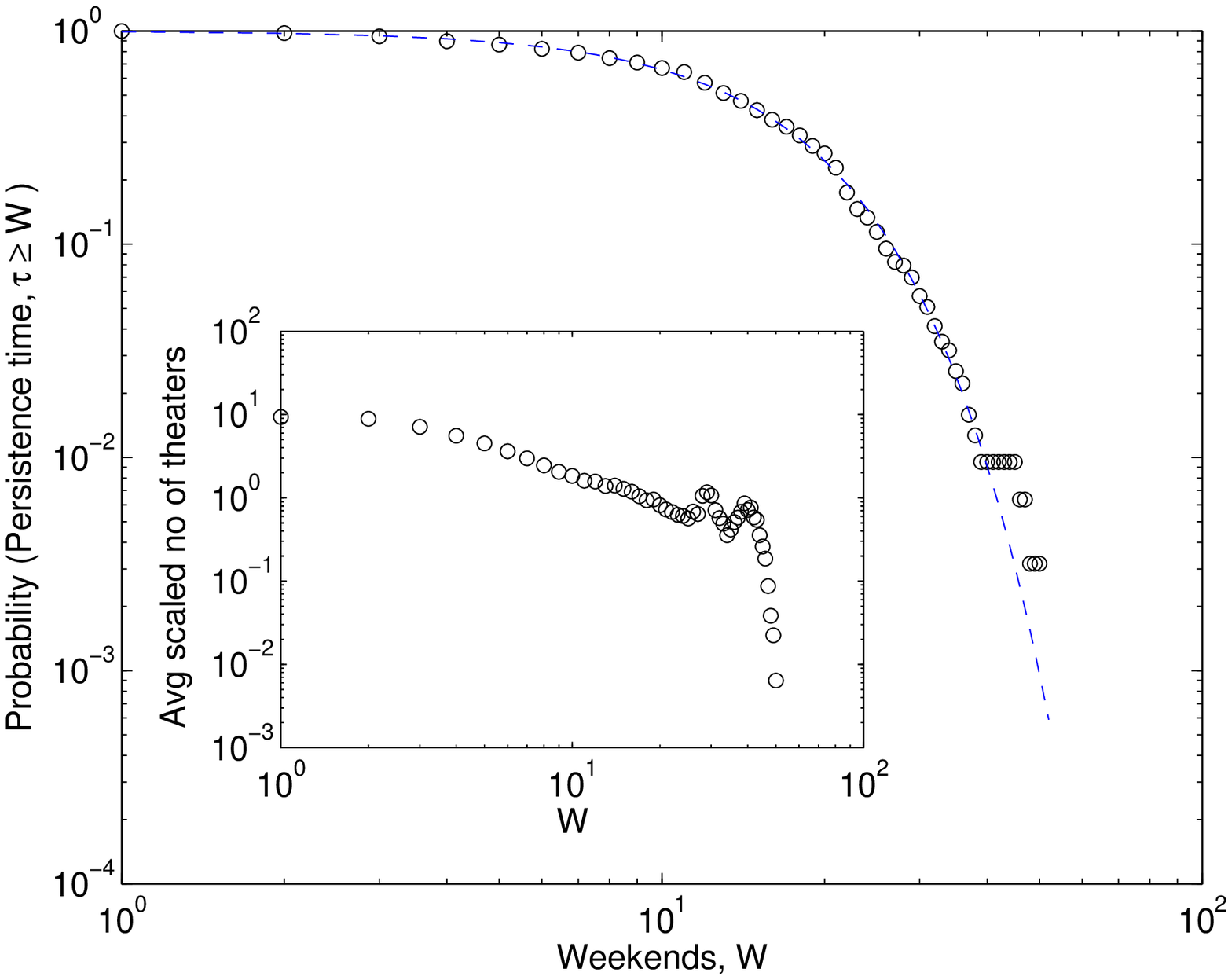}
\includegraphics[width=0.48\linewidth,clip]{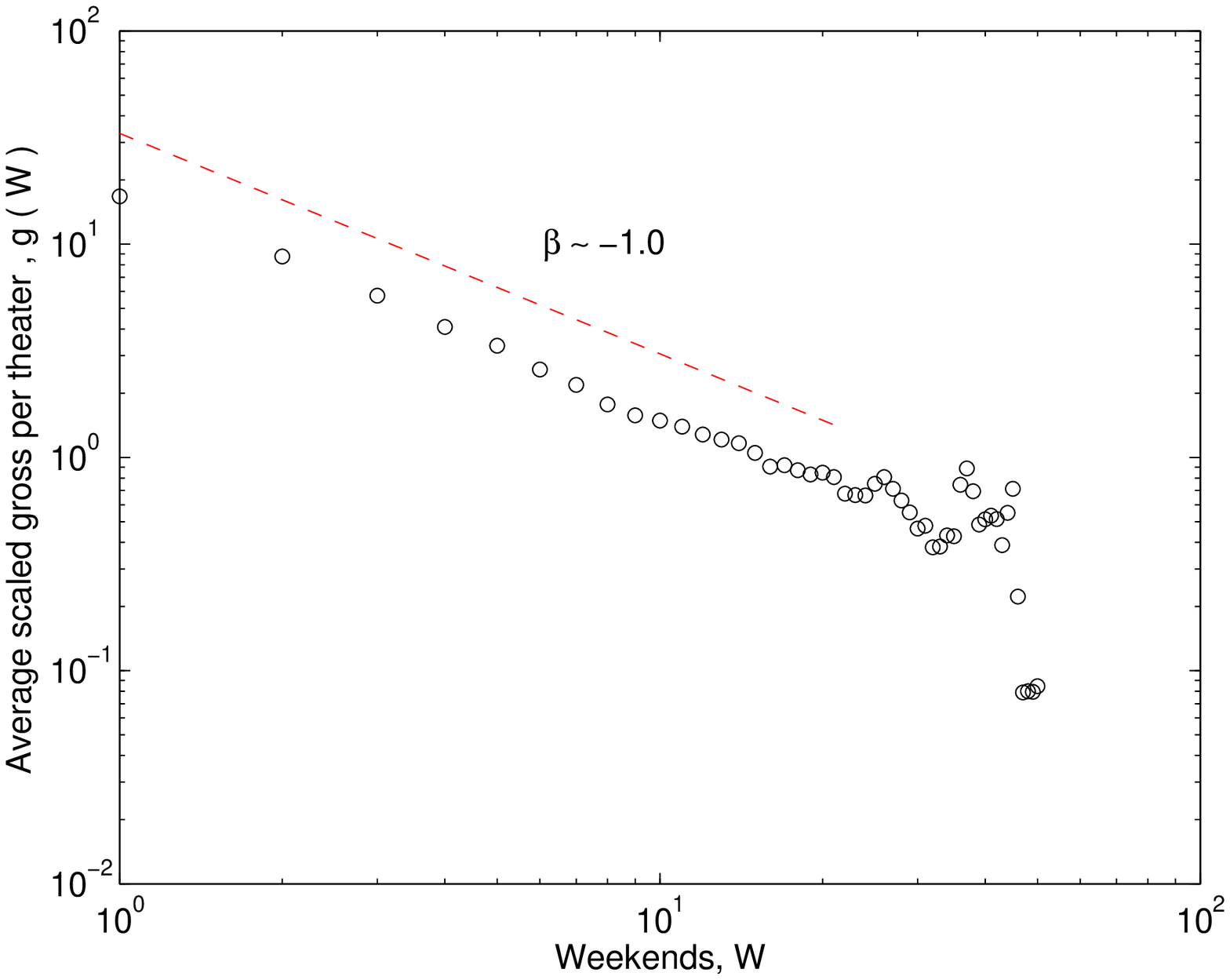}
\caption{Time evolution of movie income for all movies released across
theaters in USA in the year 2002.
(Left) Cumulative probability distribution of movie persistence time
$\tau$ (in terms of weekends). The broken line shows fit with a stretched
exponential distribution $P ( x ) = exp (-[x/x_0]^c)$, with $x_0
\simeq 16.5$ and $c \simeq 1.75$. The inset shows the number
of theaters (scaled by the average number of theaters that
a movie was shown in
its theatrical lifespan) in which a movie runs after $W$ weekends,
averaged over the number of movies that ran for that long.
(Right) Weekend gross per theater for a movie (scaled by the average weekend 
gross over its theatrical lifespan), $g ( W )$,
after it has run for $W$ weekends,
averaged over the number of movies that ran for that long.
The initial decline follows a power-law with exponent
$\beta \simeq -1$ (the fit is shown by the broken line). 
}\label{fig:3}
\vspace{-0.35cm}
\end{figure}

\noindent
To make a quantitative analysis of the relative performance of movies in 
a given year (say 2002), we define the persistence time $\tau$ of a movie 
as the
time (measured in number of weekends) upto which it is being shown at theaters.
Fig. \ref{fig:3} (left) shows that most movies run for upto about 10 weekends,
after which there is a steep drop in their survival probability. The 
tail is almost entirely composed of sleepers, the best performance being
by {\em My Big Fat Greek Wedding} ($\tau = 51$ weekends). The
inset shows the time-evolution of the average 
number of theaters showing a movie.
It suggests an initial power-law decay followed by an exponential
cut-off. We also look at the time-evolution of the gross per theater, $g$.
This is a better measure of movie popularity, because a movie that is being
shown in a large number of
theaters has a bigger income simply on account of higher
accessibility for the 
potential audience. Unlike the overall gross that decays exponentially 
with time, the gross
per theater shows a power-law decay with exponent $\beta \simeq -1$
[Fig. \ref{fig:3} (right)]. 

\vspace{-0.5cm}
\section{Conclusions}
\vspace{-0.25cm}
To conclude, we have shown that movie income distribution has a power-law 
tail with Pareto exponent $\alpha \simeq 2$. This is suggestive of a 
possible universal exponent for many 
popularity distributions. The exponent is identical for the opening as 
as well as the total gross distribution. Since the Pareto tail appears
at the opening week itself, it is unlikely that the mechanism for
generating this behavior involves information exchange between moviegoers.
Also, as mentioned before, conventional
asset exchange models don't apply in this 
case. Therefore, explaining the
Pareto tail of the income distribution, as well as the distribution of
the time-evolution of movie income, is an interesting challenge to
theories of distributions with power-law tails.

\vspace{0.1cm}
\noindent
{\small We would like to acknowledge helpful discussions
with S. Raghavendra, S. S. Manna, D. Stauffer, P. Richmond and 
B. K. Chakrabarti.}

%
%

%
%

\printindex
\end{document}